\documentstyle[prd,aps,psfig]{revtex}
\draft
\begin{document}
\twocolumn
\title{
Magnetically Induced "Dry" Water  Like Structure of Charged Fluid
at the Core of a Magnetar
}
\author{
 Sutapa Ghosh$^{a)}${\thanks{E-Mail: sutapa@klyuniv.ernet.in}}
 and Somenath
 Chakrabarty$^{a),b)}${\thanks{E-Mail: somenath@klyuniv.ernet.in}}
}
\address{
$^{a)}$Department of Physics, University of Kalyani, Kalyani 741 235,
India and
$^{b)}$Inter-University Centre for Astronomy and Astrophysics, Post Bag 4,
Ganeshkhind, Pune 411 007, India
}
\date{\today}
\maketitle
 
\begin{abstract}
It is shown that charged fluid, e.g., electron gas or proton matter at
the core of a magnetar exhibit super-fluid (frictionless) like property if the 
magnetic field strength is high enough to populate only the zeroth Landau 
levels.
\end{abstract}

One of the oldest subject- {\sl{the effect of strong magnetic field
on dense charged particle system}} has got a fresh life with the
observational discovery of a few magnetars. These exotic objects are
assumed to be the strongly magnetized young neutron stars and also the
possible sources of soft gamma repeaters (SGR) and anomalous X-ray
pulsars (AXP) \cite{R1,R2,R3,R4}. During the past few years a lot of
work have been done on
the effect of strong magnetic field on various physical properties of
dense astrophysical matter \cite{R5,R6,R7,R8,R9}. An intensive studies 
have also been
done on the structural deformation of neutron stars by strong magnetic
field, emission of gravity waves from such rotating deformed objects etc
\cite{R10,R11,R12,R13,R14}.

From the observational data of SGR and AXP, the strength of surface
magnetic field of magnetars are predicted to be $\geq 10^{15}$G. Then 
by scalar Virial theorem it is very easy to show that the magnetic field 
strength
at the core region may go up to $10^{18}$G. This is of course strong
enough to affect most of the physical processes taking place at the core
region of magnetars. In particular, the physical properties like
equation of states of dense hadronic matter, electromagnetic and weak 
processes, quark-hadron
phase transition, transport properties etc. should change significantly in
presence of such strong magnetic field \cite{R5,R6,R7,R8,R9}. Since the 
hadronic matter is
assumed to be in $\beta$-equilibrium, in presence of a strong magnetic
field, the properties of neutron sector will also change significantly. 
It is known that in the
relativistic limit, if the cyclotron quantum of the $i$th charged species
exceeds the rest mass of the constituent, the quantum mechanical effect of
strong magnetic field on the $i$th charged component becomes
significant. In other wards, the Landau levels of the $i$th charged
species are populated beyond this critical field strength. 
This physical phenomenon is called the {\sl{Landau
diamagnetism}} \cite{R15}. The critical value of magnetic field strength for
the $i$th charged component is given by $qB^{(c)(e)}=m_i^2$, where $q$ is
the magnitude of charge carried by the particle and $m_i$ is the rest
mass of the particle (we have assumed $\hbar =c=k_B=1$). For electron of rest
mass $m_e=0.5$MeV, this critical strength is $B^{(c)(e)} \approx 4.4\times
10^{14}$G. In the case of proton, this typical value
changes to $B^{(c)(e)}m_p^2/m_e^2$, which is extremely high to achieve
even at the core region of a magnetar with extremely strong magnetic
field. Therefore, the quantum mechanical
effect of strong magnetic field may be neglected in the case of protons.
However, if the kinetic pressure of proton matter in presence of strong
external magnetic field is compared
with the corresponding kinetic pressure of magnetically affected
electron gas, the proton pressure exceeds the electronic contribution
under charge neutrality condition and in $\beta$-equilibrium (with
neutron). This is obviously unphysical, since protons are much more
massive than electrons. To remove this anomaly, we assume, that protons
are also affected quantum mechanically by strong magnetic field. We
further consider the simplest physical picture of $n-p-e$ system in
$\beta$-equilibrium and also assume that $p-e$ system is charge neutral.
Now it is generally believed that under such extreme physical condition
at the core region of a neutron star the neutron matter exhibits
super-fluidity, whereas the proton sector becomes super-conducting
\cite{R16}.
However, in the present article, we are
not interested on the type of super-conductivity of proton matter.
Since the kinetic energy of electrons are a few orders of
magnitude larger than the super-conducting band gap obtained from BCS
theory, they never show
super-conducting behavior. Now following the relativistic theory of
super-fluidity and super-conductivity of Fermionic system, developed by 
Bailin and Love \cite{R17}, we
have estimated the critical strength of magnetic field at which the
super-conductivity of proton matter is completely destroyed. The typical
value is $\sim 10^{16}$G corresponding to the density and temperature
relevant for neutron star core. Therefore, at the core region of a magnetar
the possibility of proton super-conductivity may be ruled out. However,
the super-fluidity of neutron matter remains unaffected.

In this article we shall try to show from the study of relativistic 
transport theory of dense electron gas or proton matter, that in 
presence of an ultra-strong magnetic field, the shear
viscosity coefficient
vanishes or becomes negligibly small. In fact, when only
the zeroth Landau levels are occupied by the charged particles, the
shear viscosity coefficient vanishes
identically. In fig.(1) we have shown the curve ($B$-$n_B$ diagram) that
separates the region where only zeroth Landau levels are populated
(indicated by "zero" in the figure) and
the region where other non-zero Landau levels are also occupied by the
electrons (indicated by "non-zero" in the figure). The curve corresponding
to proton matter exactly coincides with the curve for electron gas. We
have further noticed that  nature of the curve is insensitive with
the increase in temperature as long as it is $<40$MeV. As the
temperature increases further, the curve simply goes up, i.e., at higher
temperature much
stronger magnetic field is needed to populate only the zeroth Landau levels.
Therefore, if one crosses the curve from below, enters the zero Landau
level zone both for electron and proton. This curve also indicates that
in the region "zero" the charged fluid is frictionless (zero shear
viscosity). However, the charged fluid, in particular the proton matter
is not super-conducting in this region. The electrical conductivities of
both electrons gas and proton matter remain finite in this zone.

Now to show that the shear viscosity coefficient of charged particle
system vanishes in presence of  extremely strong magnetic field or in
other wards, in the "zero" region as shown in the figure, at which
only the zeroth Landau levels are occupied, we start
with the conventional form of relativistic Boltzmann equation following
de Groot \cite{R18}. The general form of relativistic Boltzmann equation is
given by
\begin{equation}
[p^\mu\partial_\mu +\Gamma_{\nu \lambda}^\mu p^\nu p^\lambda
\partial_{p_\mu} ]f(x,p)=C[f]
\end{equation}
where $f(x,p)$ is the distribution function for electron or proton and
$C[f]$ is the collision term which contains the rates of all possible
elementary processes. The second term on the left hand side is
the external force field term. In the case of flat space-time geometry
or in absence of an external field this term becomes zero. This particular 
term also does not play any  direct
significant role in the evaluation of  viscosity coefficient.

To obtain an expression for the shear viscosity coefficient of charged
fluid, e.g., electron gas or proton matter, we follow the theoretical
technique described in the famous book on Relativistic Kinetic Theory by
de Groot \cite{R18}. We replace collision term by the relaxation time
approximation, given by
\begin{equation}
C[f]=-\frac{p^0}{\tau} (f(x,p)-f^0(p))
\end{equation}
where $\tau$ is the relaxation time, which again depends on the rates of 
all the
relevant (strong, electromagnetic and weak) processes and $f^0(p)$ is
the equilibrium distribution function (in this case it is the Fermi 
distribution
of either electron or proton). Let us consider a fluid element moving
with the hydrodynamic
four velocity $u^\mu$, then we can define a second rank tensor 
$\Delta^{\mu \nu}=
g^{\mu \nu} -u^\mu u^\nu$ which projects the space-like component of any
general four vector, say $a^\mu$. Where $g^{\mu\nu}={\rm{diag}}(1~ -1~ -1~
-1) $ is the metric tensor. Hence it is possible to define the convective time
derivative $D=u^\mu\partial_\mu$ and gradient operator $\nabla^\mu
=\Delta^{\mu \nu}\partial_\nu$. The zeroth order dynamical equations 
for electron gas or for the proton matter are then obtained from the
stability condition of Boltzmann equation and the first order
decomposition of $(Df_k)^{(1)}$, given by (see ref. \cite{R18})
\begin{equation}
(Df_k)^{(1)}=\frac{\partial f_k^0}{\partial n_k} (Dn_k)^{(1)}
+\frac{\partial f_k^0}{\partial T} (DT)^{(1)}
+\frac{\partial f_k^0}{\partial u^\mu} (Du^\mu)^{(1)}
\end{equation}
where $k$ represents either electron or proton.
The stability equations are obtained from the conservation laws of 
charge current and energy momentum tensor 
Then we have for the electron gas
\begin{eqnarray}
Dn&=&-n\nabla_\mu u^\mu \nonumber \\
Du^\mu&=& \frac{1}{nh}\nabla^\mu P\nonumber \\
C_vDT&=&F(\mu_e, T)\nabla_\mu u^\mu
\end{eqnarray}
Similar sets of equation can also be obtained for proton matter.
To obtain the set of dynamical equations (eqn.(4)) for either electron gas or
proton matter, we have assumed that $n-p-e$ system is a reactive mixture
and followed the theoretical techniques as discussed in ref. \cite{R18}
for such a system. In the
above equations, $n$ is the number density, $P$ is the kinetic
pressure, $h=\varepsilon +Pn^{-1}$ is the enthalpy per electron,
$\varepsilon$ is the average energy per particle, $C_v=\partial
\varepsilon/\partial T$ is the specific heat per particle at constant
volume  for the electron gas or proton matter 
and $F(\mu_i,T)=P(\mu_i,T)/n(\mu_i,T)$. The function $F(\mu_i,T)=T$ in
the classical case, here $i=e$ or $p$.

It is further assumed that the system is very close to its equilibrium
configuration. Then we can linearize the transport equation by the
ansatz
\begin{equation}
f(x,p)=f^0(p)(1+\chi(x,p))
\end{equation}
where $\chi(x,p)$ is the first order deviation from equilibrium
configuration.
Now the standard technique which is generally followed to obtain the transport
coefficients of electron gas or of proton matter is to linearize Boltzmann 
equation using the above ansatz
and decompose partial derivative into convective time derivative and
gradient operator and finally use the dynamical equations (eqn.(4)) and the
relativistic version of Gibbs-Duhem equation, given by
\begin{equation}
\frac{1}{n}\nabla^\mu P=h\frac{\nabla^\mu T}{T}+T\nabla^\mu \left (
\frac{\mu_e}{T}\right )
\end{equation}
where $\mu_e$ is the electron chemical potential.
Combining all these, we finally  get
\begin{equation}
\chi(x,p)=\frac{\tau (1-f^0)}{p^0T}(QX+p^\mu p^\nu \bar X_{\mu \nu} -
p_\nu(p^\mu u_\mu -h)X_q^\nu)
\end{equation}
where $X$, $\bar X_{\mu \nu}$ and $X_q^\nu$ are the driving forces for
volume viscosity, shear viscosity and heat conduction respectively. In
the present article we are only interested on the driving force for
shear viscosity, given by
\begin{equation}
\bar X_{\mu \nu}=(\nabla_\mu u_\nu -\frac{1}{3}\Delta_{\mu
\nu}\nabla_\sigma u^\sigma)
\end{equation}
The
next step is to substitute the linearized form of distribution function
in the expression for energy momentum tensor, given by
\begin{equation}
T^{\mu \nu}= \sum_{\nu=0}^\infty
\int \frac{d^3p}{p^0(2\pi)^3}p^\mu p^\nu f(x,p)
\end{equation}
In this case the phase space volume element $d^3p=dp_zeB/(2\pi^2)$ and $\nu$
is the Landau quantum number.
Substituting the linearized form of $f(x,p)$, we get
\begin{equation}
T^{\mu \nu}= T_{\rm{eq}}^{\mu \nu}+ T_{\rm{non}}^{\mu \nu}
\end{equation}
where
\begin{equation}
T_{\rm{eq}}^{\mu \nu}=\epsilon u^\mu u^\nu  -P\Delta^{\mu \nu}
\end{equation}
and
\begin{equation}
T_{\rm{non}}^{\mu \nu} = \sum_{\nu=0}^\infty \int \frac{d^3p}{p^0(2\pi)^3}
p^\mu p^\nu 
\chi(x,p) f^0(p)
\end{equation}
are respectively the equilibrium and non-equilibrium parts of energy
momentum tensor.
To obtain an expression for shear viscosity coefficient of charged fluid, 
we consider a
special kind of flow. We assume that the magnetic field is along
the positive $z$-direction and the flow is also in the same direction. Then
we have 
\begin{equation}
T_{\rm{non}}=-\eta \frac{\partial u_z}{\partial r}
\end{equation}
where $\eta$ is the coefficient of shear viscosity. Substituting the
expression for $\chi$ in the non-equilibrium part of energy momentum
tensor, putting $\mu=r$ and $\nu=z$, and finally equating the
coefficient of the driving force $\partial u_z/\partial r$, we get 
\begin{equation}
\eta= \frac{1}{T} \sum_{\nu=0}^\infty \int \frac{d^3p}{p^0(2\pi)^3}\tau(p_z)
(p^r p^z)^2
f^0(p)(1-f^0(p)
\end{equation}

Let us first consider the electron gas. In
this case if the magnetic field strength is greater than the
critical value $B^{(c)(e)}$, the modified form of single particle energy is 
given by
$\varepsilon_\nu=(p_z^2+m_e^2+2\nu eB)^{1/2}$ and the transverse
momentum $p_r=(2\nu eB)^{1/2}$,
where $\nu$ is the Landau quantum number. It is known that with the
increase of magnetic field strength the maximum value  $\nu_{\rm{max}}$ up to
which the Landau levels are occupied by electrons, decreases and in the
limiting case, when $B$ becomes extremely high, $\nu_{\rm{max}}$ becomes
zero. As indicated in fig.1, it also depends on the density of electron
gas. In this extreme physical scenario the transverse part of electron momentum
$p_r$ also vanishes and as a consequence the shear viscosity coefficient
of electron gas (see eqn.(14)) also becomes zero. The same is also true for the 
proton matter. 

Therefore in the extreme physical situation, when the magnetic field
strength is sufficiently high, the viscosity coefficient of charged
fluid at the core region of a magnetar vanishes, which means the matter
becomes frictionless, i.e., the charged fluid, including the electron
gas behaves "like" super-fluid. Unlike the dense hadronic matter at the core 
region of a
normal neutron star, where (i) neutron matter shows
super-fluidity, (ii) proton matter also exhibits super-fluid "like"
property, but it does not show super-conductivity and (iii) the electron
gas also shows super-fluid "like" behavior. Therefore all the constituents at
the core of a magnetar in this simplified picture behave like
frictionless fluid without the super-conducting properties of proton
matter. 

Because of the presence of dissipative processes at the core region of a
magnetar, entropy will be
produced through irreversible processes. The entropy production
$\sigma$ is defined as
\begin{equation}
\sigma=\partial_\mu s^\mu
\end{equation}
where $s^\mu$ is the entropy current. It can be shown that, in the first
Chapman-Enskog approximation, the entropy production because of shear
viscosity only is given by
\begin{equation}
\sigma =\frac{1}{T} \bar \Pi^{\mu \nu}\bar X^{\mu \nu}
\end{equation}
where $\bar \Pi^{\mu \nu}$ is the traceless part of viscous pressure
tensor. Since the matter becomes frictionless (non-viscous) in presence
of ultra strong magnetic field, the entropy production simply becomes
zero and right hand side of eqn.(16) vanishes. Since the elementary
processes taking place at the core region are not stopped by the strong
magnetic field, a chemical evolution of the matter is
possible in this extreme physical condition without the  generation of
entropy. In reality, the entropy will be produced through other
irreversible processes, e.g., volume viscosity and heat conduction which 
are non-zero in such exotic matter.
This is of course very very strange behavior of charged fluid.
The charged fluid behaves like "dry" water if only the
zeroth Landau levels are populated in presence of strong quantizing
magnetic fields.
\begin{figure}
\psfig{figure=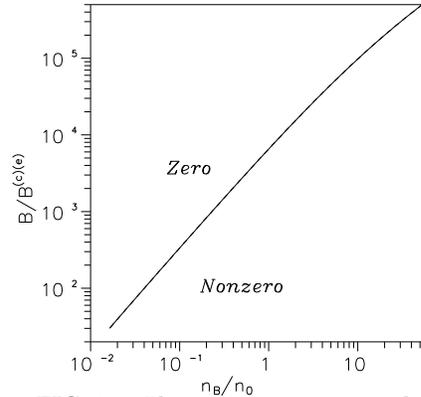,height=0.6\linewidth}
\caption{
The curve separating  only zero Landau level occupied region and
non-zero Landau level occupied region. 
}
\end{figure}

\noindent Acknowledgment: SC is thankful to the Department of Science and 
Technology, Govt. of India, for partial financial support of this work, 
Sanction number:SP/S2/K3/97(PRU).  
\end{document}